\documentclass[twoside,12pt]{article}
\usepackage{graphicx}

\def\Journal#1#2#3#4{{#1} {#2} (#4) #3 }
\def\NCA{{\em Nuovo Cimento} A}

\def\NPA{{\em Nucl. Phys.} A}

\def\PLB{{\em Phys. Lett.} B}

\def\PRL{\em Phys. Rev. Lett.}

\def\PREP{\em Phys. Rep.}

\def\PRD{{\em Phys. Rev.} D}
\def\PRC{{\em Phys. Rev.} C}

\newcommand{\be}{\begin{equation}}
\newcommand{\ee}{\end{equation}}
\newcommand{\bea}{\begin{eqnarray}}
\newcommand{\eea}{\end{eqnarray}}

\begin{document}

\title{ \vspace{1cm} Chiral Partners and their Electromagnetic Radiation \\
{\Large (Ingredients for a systematic in-medium calculation)}}

\author{S.\ Leupold,$^{1}$ M.F.M.\ Lutz,$^{2}$ M.\ Wagner$^3$
\\
$^1$ Johann Wolfgang Goethe-Universit\"at Frankfurt, Germany \\
$^2$ GSI Darmstadt, Germany\\
$^3$ Institut f\"ur Theoretische Physik, Universit\"at Giessen, Germany
}
\maketitle
\begin{abstract} 
It is argued that the chiral partners of the lowest-lying hadrons are hadronic molecules
and not three-quark or quark-antiquark states, respectively. As an example the case 
of $a_1$ as the chiral partner of the $\rho$ is discussed. Deconfinement --- or as a
precursor large in-medium widths for hadronic states --- is proposed as a natural
way to accommodate for the fact that at chiral restoration the respective in-medium 
spectra of chiral partners must become degenerate. Ingredients for a systematic
and self-consistent in-medium calculation are presented with special emphasis on
vector-meson dominance which emerges from a recently proposed systematic counting
scheme for the mesonic sector including pseudoscalar and vector mesons as active
degrees of freedom.
\end{abstract}

\section{Chiral partners}
\label{sec:chipart}

\begin{figure}
\centerline{
    \includegraphics[keepaspectratio,width=0.49\textwidth]{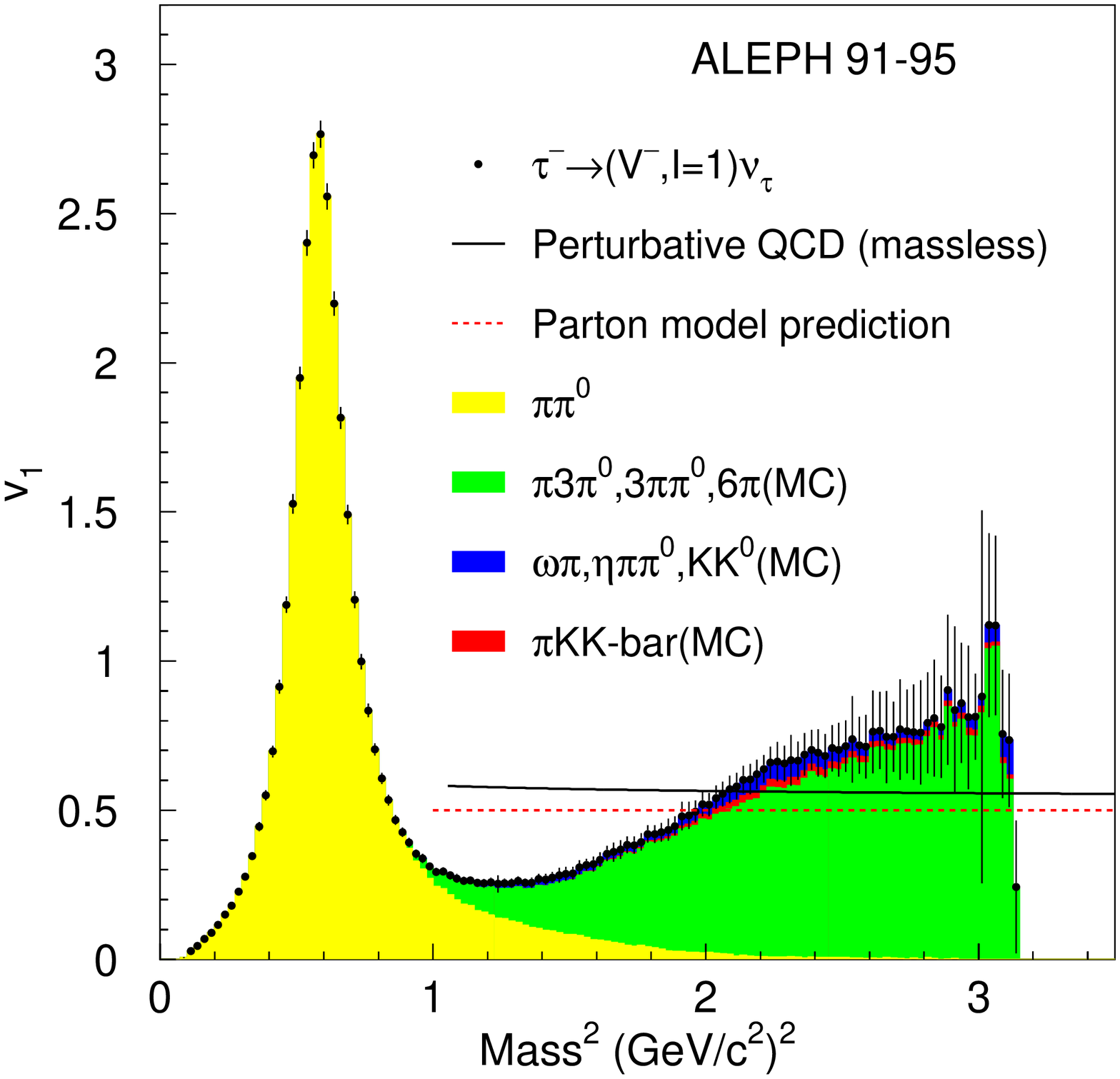}  
    \hfill
    \includegraphics[keepaspectratio,width=0.49\textwidth]{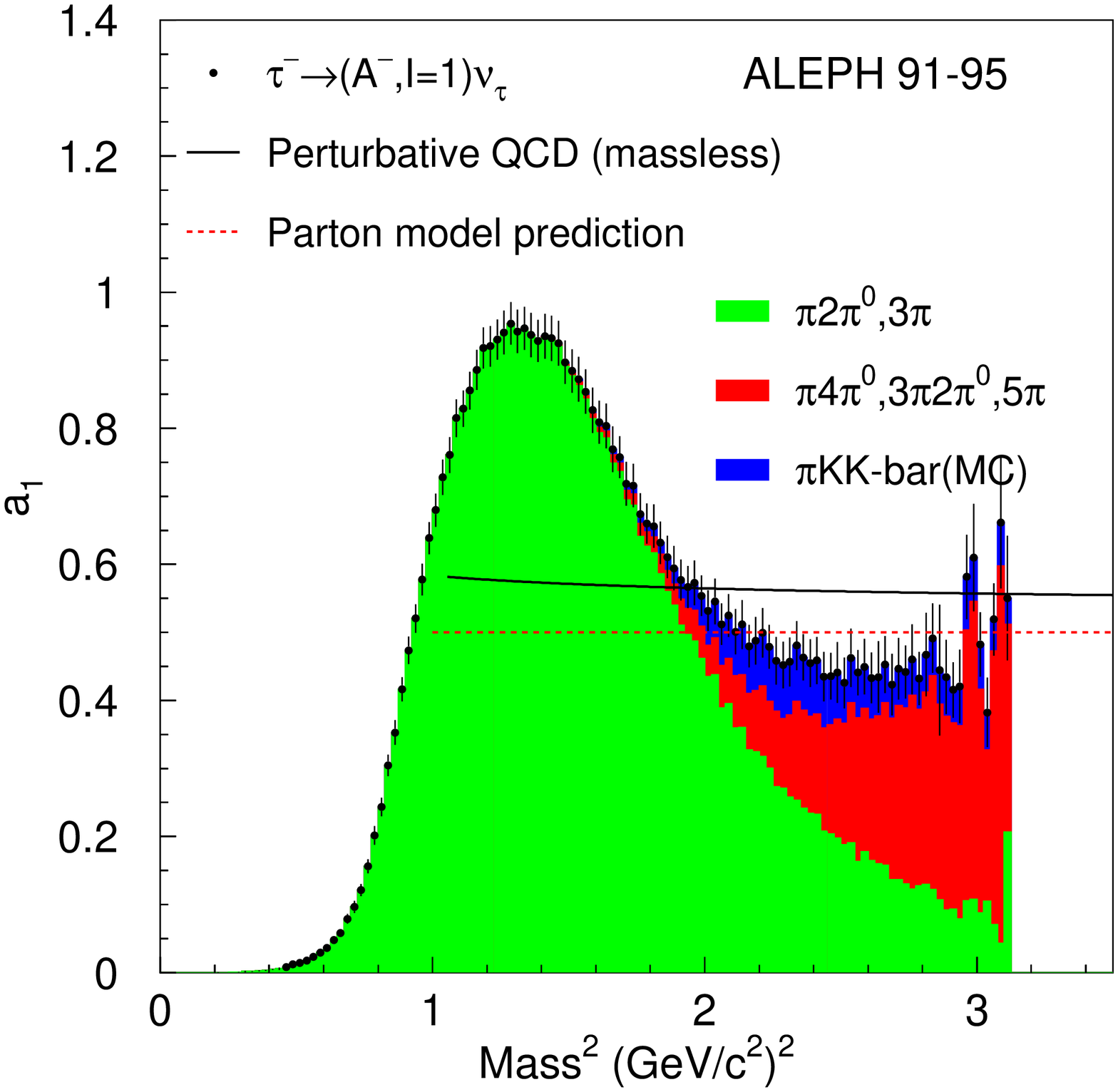}  
}
\caption{Spectral information of the vector ({\it left}) and axial-vector ({\it right}) 
current. Figs.\ taken from \cite{aleph}.} 
\label{fig:chibr-exp}
\end{figure}

Chiral symmetry breaking and its restoration in a strongly interacting medium is
one of the key issues of in-medium hadron physics, at least for the states made out
of light ($u$, $d$, $s$) quarks (see e.g.\ the talk of T.\ Hatsuda in the present 
proceedings volume). One of the clearest signs that
chiral symmetry is indeed spontaneously broken comes from a comparison of the spectra
of quark currents related by a chiral transformation, namely the vector--isovector
current $\vec j_V^\mu = \bar q \vec \tau \gamma^\mu q$
and the axial-vector--isovector 
current $\vec j_A^\mu = \bar q \vec \tau \gamma_5 \gamma^\mu q$.\footnote{Here 
$\vec\tau$ denotes the isospin matrices. A generalization to flavor SU(3) is 
straightforward.} If chiral symmetry was realized in the same way as, say, isospin
symmetry, then the spectra of the respective current-current correlators would
be (approximately) the same. The experimental results for these spectra are shown
in Fig.\ \ref{fig:chibr-exp}.
Obviously the spectra are not identical, not even approximately. 
In particular, the vector spectrum (Fig.\ \ref{fig:chibr-exp}, left) shows a peak 
below 1 GeV, the $\rho$ meson, whereas the axial-vector 
spectrum (Fig.\ \ref{fig:chibr-exp}, right) does not show
any structure below 1 GeV, but instead a broad bump at around 1.2 GeV, the $a_1$ meson.
Since the vector and the axial-vector current are related by a chiral transformation
one can call these quark currents chiral partners at the fundamental level. 
It is suggestive to call $\rho$ and $a_1$ chiral partners at the hadronic level since
they couple to the respective quark currents as seen in Fig.\ \ref{fig:chibr-exp}. 
Obviously, due to chiral symmetry breaking the masses of $\rho$ and $a_1$ are not the
same. In the following we shall show strong indications that chiral partners are even
different in nature.
To which extent the phrase ``chiral partners'' is pure semantics or contains physics
is discussed in more detail in \cite{proc-meson08}.

\section{Nature of chiral partners}
\label{sec:nat}

We start with the lowest-lying hadronic states (in flavor SU(3)), 
the nucleon octet, the pion nonet, the $\Delta$ decuplet and the $\rho$ nonet.
In the following we assign to these states the label ``LLH'' (= lowest-lying hadrons).
Without much doubt these states are dominantly quark-antiquark or three-quark states, 
respectively (concerning the $\rho$ meson see e.g.\ \cite{proc-meson08}).
On the other hand, the chiral partners of the LLH states can be understood as
dynamically generated states, i.e.\ in a somewhat oversimplified language as
hadronic ``molecules''. For the $N^*(1535)$, the chiral partner of the nucleon,
it has been demonstrated in \cite{kaiser} and many follow-up works that it emerges
from the coupled-channel dynamics of $\eta N$, $K \Lambda$, \ldots.
Many works have been devoted to the $\sigma$ meson, the chiral partner of the pion.
For example in \cite{oller-oset} the $\sigma$ emerges as a dynamically generated
state in pion-pion scattering. In \cite{lutz-Delta} it has been argued that the
$\Delta^*(1700)$, the $N^*(1520)$ and their respective flavor partners, 
which can be seen as the 
chiral partners of the $\Delta$ decuplet, are hadronic coupled-channel ``molecules''.
Finally the $a_1$ multiplet is generated dynamically in \cite{lutz-a1}.
We note in passing that also the $b_1$ multiplet can be viewed as the chiral
partner of the $\rho$ multiplet \cite{caldi}. This apparent ambiguity is resolved
in the sense that also the $b_1$ multiplet is generated on equal 
footing in \cite{lutz-a1}. 

\begin{figure}[t]
  \begin{minipage}[b]{0.49\textwidth}
    \begin{center}
      \includegraphics[keepaspectratio,width=0.8\textwidth]{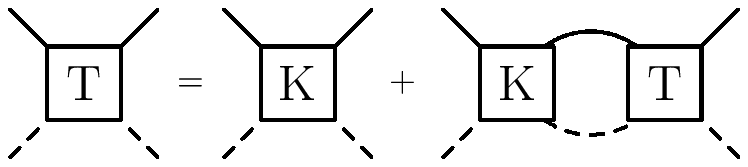} \\[3em]
      \includegraphics[keepaspectratio,width=0.48\textwidth]{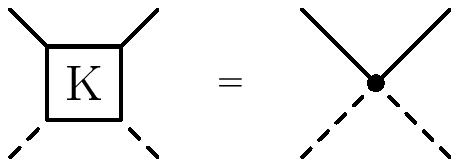}  
    \end{center}  
  \end{minipage}
  \hfill
  \begin{minipage}[b]{0.45\textwidth}
    \includegraphics[keepaspectratio,width=\textwidth]{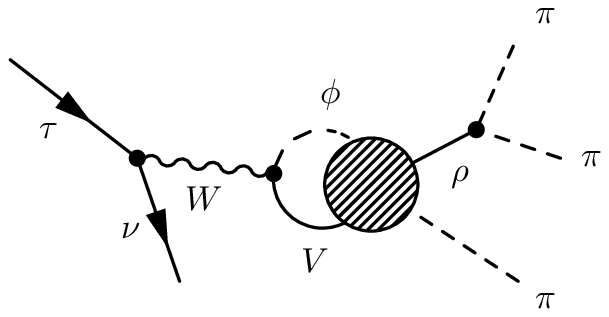}
  \end{minipage}
  \caption{{\it Left, top:} Generic Bethe-Salpeter equation for the scattering of a 
    Goldstone boson (dashed lines) on an LLH state (full lines). 
    {\it Left, bottom:} In the framework described in the main text the kernel K
    of the Bethe-Salpeter equation is the chiral lowest-order interaction, the
    Weinberg-Tomozawa point interaction.
    {\it Right:} Description of the decay $\tau \to \nu_\tau + 3\pi$ with the $a_1$ as
    a final-state interaction effect. The blob denotes the S matrix for the scattering
    of pseudoscalar states $\phi$ on vector states $V$. See main text for details.}
  \label{fig:diagr-tau}
\end{figure}

The works cited above essentially use the same framework for dynamical generation:
One studies the scattering of an LLH state on Goldstone bosons for the channel of 
interest, i.e.\ the one with the quantum numbers of the chiral partner of the LLH state.
The scattering matrix $T$ is determined from the Bethe-Salpeter equation as shown on the 
left-hand side of Fig.\ \ref{fig:diagr-tau}. The input for the Bethe-Salpeter equation,
the interaction kernel, is always of the same type: One considers the lowest order
of the chiral interaction, the Weinberg-Tomozawa point interaction \cite{WT}. 
Consequently, due to chiral symmetry breaking, the strength
of this interaction is fixed model independently $\sim F_\pi^{-2}$, where $F_\pi$
denotes the pion-decay constant. The Bethe-Salpeter equation requires renormalization
to obtain a well-defined meaning. As shown in \cite{lutz-a1,hyodo} the renormalization
point for the loop appearing in the Bethe-Salpeter equation is actually fixed, e.g.\
by requiring approximate crossing symmetry for the scattering 
matrix $T$ \cite{lutz-Delta,lutz-a1}. Thus, there are no free parameters for the
calculation of the scattering amplitude using the leading order chiral interaction
as an input. Peaks in the scattering amplitude signal the appearance of dynamically
generated resonances. 

\begin{figure}[ht]
  \includegraphics[keepaspectratio,width=0.49\textwidth]{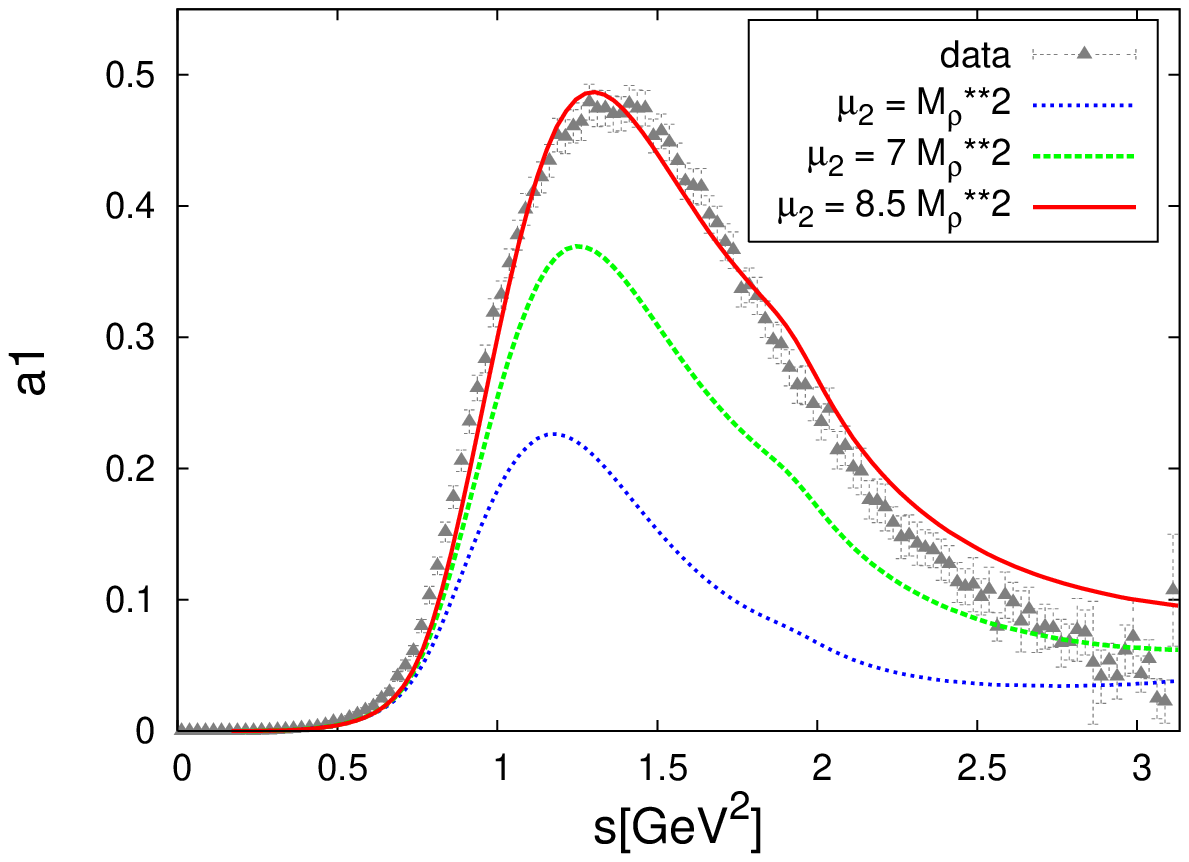}  
  \hfill
  \includegraphics[keepaspectratio,width=0.49\textwidth]{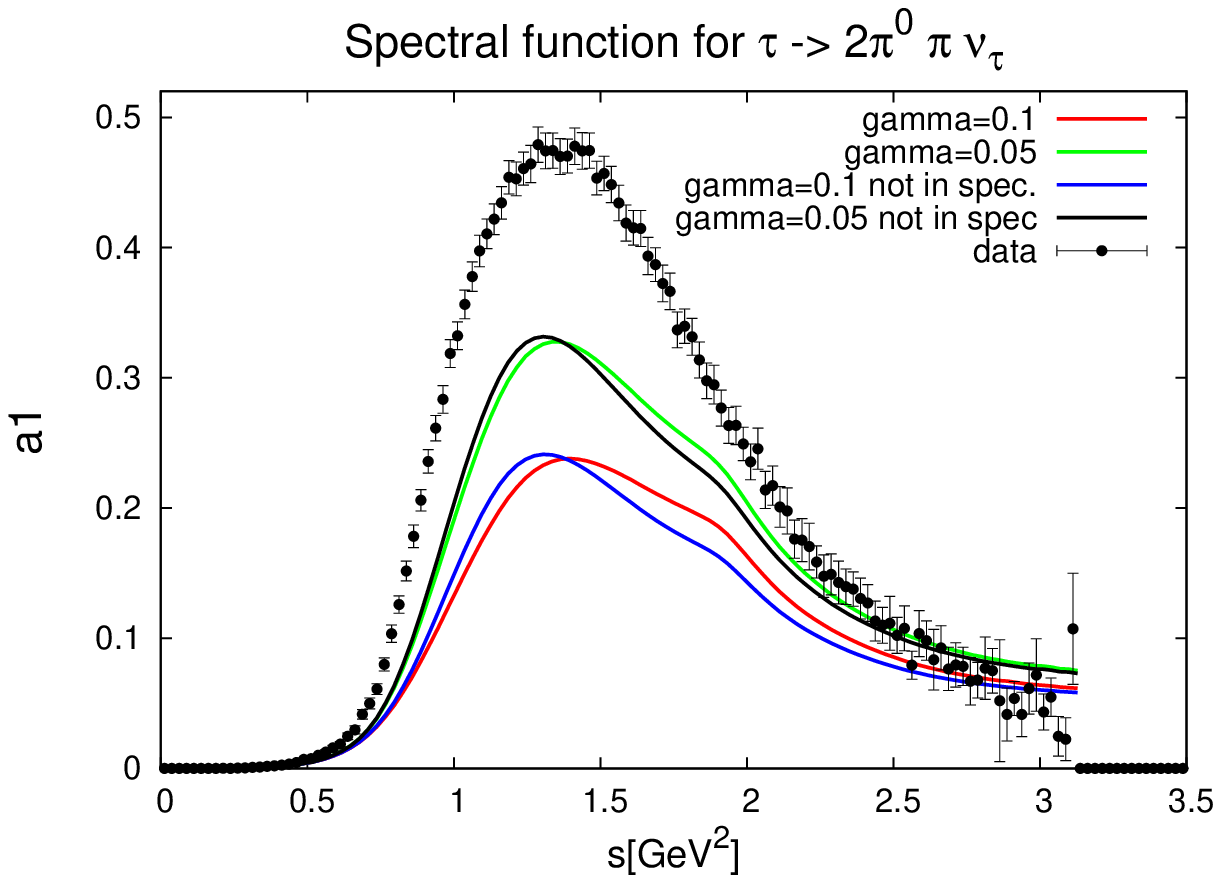}  
  \caption{{\it Left:} Low-energy, i.e.\ 3-pion spectral information, of the axial-vector
    current in the scenario where the $a_1$ is dynamically generated as compared
    to data. Figure taken from \cite{wagner-long}. 
    {\it Right:} Corresponding in-medium spectrum from a simple model. Figure taken
    from \cite{proc-meson08}. See main text for details.}
  \label{fig:a1-plots}
\end{figure}

In the following we concentrate on one specific channel, namely on
(the low-energy part of) the axial-vector spectrum shown 
in Fig.\ \ref{fig:chibr-exp}, right. Not shown is the fact that there is more 
differential information available, namely Dalitz plots for the three-pion
hadronic final state. These Dalitz plots show that the three-pion final state
is correlated to a $\pi$-$\rho$ state \cite{wagner-long}. 
Above we have described the scenario where the chiral partners of the LLH states
are dynamically generated. For the case at hand this implies that the two-body
state of vector meson and Goldstone boson is subject to a strong final-state 
interaction which creates the $a_1$ bump seen in Fig.\ \ref{fig:chibr-exp}, right.
We shall study in the following how well this scenario works. The corresponding processes
are depicted in Fig.\ \ref{fig:diagr-tau}: The right panel shows
the whole process from which the experimental information is extracted, the
decay $ \tau \to \nu + 3\pi$. From the weak-interaction vertex the hadronic two-body
state of vector meson and Goldstone boson emerges ($\rho$-$\pi$ and $K^*$-$K$).
Its final-state interaction is obtained from the Bethe-Salpeter 
equation \cite{lutz-a1,oset-a1} 
shown on the left-hand side of Fig.\ \ref{fig:diagr-tau} and described above.
There is one parameter not fixed by the general considerations: the renormalization
point, $\mu_2$, of the entrance loop for the rescattering process, 
i.e.\ the loop explicitly displayed in Fig.\ \ref{fig:diagr-tau}, right. 
Essentially it renormalizes the $W$-to-hadrons 
vertex. 
We keep the renormalization point $\mu_2$ as a free
parameter and study in Fig.\ \ref{fig:a1-plots}, left, how the results 
depend on it \cite{wagner-long}. Also shown in this plot are the three-pion 
final-state data from \cite{aleph}. 
Obviously, the variation of the only free parameter $\mu_2$ changes height and width
of the result, but not so much the peak position of the dynamically generated $a_1$ 
state.
In addition, one sees that a good agreement with the data (peak position, height and
width) is obtained with only one free parameter \cite{wagner-long}. 
This success supports the scenario
of dynamical generation of the chiral partners of the LLH states.

\section{What happens at chiral restoration?}
\label{sec:chirest}

Typically a spontaneous symmetry breaking is lifted at some temperature and/or density. 
(For example, for a Ferro magnet the spontaneous magnetization vanishes and rotational
invariance is restored at the Curie temperature.) Consequently, the spectral 
information of the vector and the axial-vector
current become identical at the point of chiral restoration. 
There are various scenarios conceivable
how this degenerate in-medium spectrum might look like \cite{shuryak}. 
Here we briefly discuss only two.
The degeneracy scenario: 
In vacuum the $\rho$ meson is dominantly a single-particle state at the hadronic level
(and not a pion-pion correlation \cite{proc-meson08}). 
If the $\rho$ meson was still dominantly a 
single-particle state at the point of chiral restoration --- i.e.\ if it still 
shows up as a prominent peak in the spectrum ---, this would require the existence 
of another single-particle state at the hadronic level with opposite parity, 
i.e.\ an axial-vector state. Since we have shown that
the $a_1$ meson is well described as a two-particle state, a $\rho$-$\pi$ correlation, 
there should be another, i.e.\ higher-lying axial-vector state which becomes 
degenerate with the $\rho$ meson.
We cannot exclude this possibility, but would regard it as rather unnatural that in 
vacuum such a state is 
so high in mass. Within our formalism we have not much to say about this scenario.
The melting scenario: It might appear that the $\rho$ meson dissolves already in 
hadronic matter. This should be understand as a precursor to deconfinement \cite{rapp}. 
Then also the $a_1$ meson should dissolve. In principle, this can be tested in our 
approach. In the following we present a very simple model:
We increase the width of the $\rho$ meson by a constant (by 50 or 100 MeV, respectively) 
and study what happens to the dynamically generated $a_1$. It must be stressed that 
this model should be regarded as a precursor to more serious calculations. 
In particular, an in-medium width of the $\rho$ meson would
not be independent of the momentum of the $\rho$ meson relative to the 
medium \cite{post}.
In addition, one also expects a strong in-medium effect on the pion and not only on 
the $\rho$ meson (see e.g.\ \cite{post} and references therein).
These aspects are not covered by the simple model studied here. The result is shown in 
Fig.\ \ref{fig:a1-plots}, right. The upper/lower two curves correspond to an 
increase of the $\rho$ meson width by 50/100 MeV. The difference between the respective
two curves close to each other is not relevant for the present 
purpose.\footnote{For one curve all vector-meson propagators in the rescattering process
are changed, for the other only (the last) one.} 
Obviously, the $a_1$ meson also melts, 
if the $\rho$ meson melts.
This does not prove that the melting scenario is the correct approach to 
chiral restoration, but at least we obtained a consistent picture. 
In a somewhat sloppy way, one might say that the problem
of the missing partner of the $\rho$ meson on the single-particle level is solved by 
deconfinement.

\section{On vector-meson dominance}
\label{sec:vmd}

\begin{figure}[ht]
  \centering
  \begin{minipage}[c]{0.3\textwidth}
    \includegraphics[keepaspectratio,width=\textwidth]{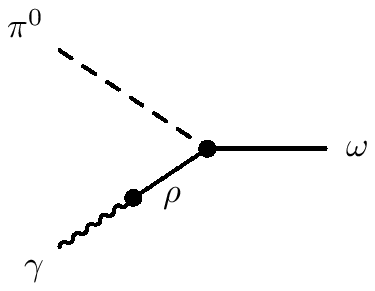}  
  \end{minipage}
  \hspace*{6em}
  \begin{minipage}[c]{0.3\textwidth}
    \includegraphics[keepaspectratio,width=\textwidth]{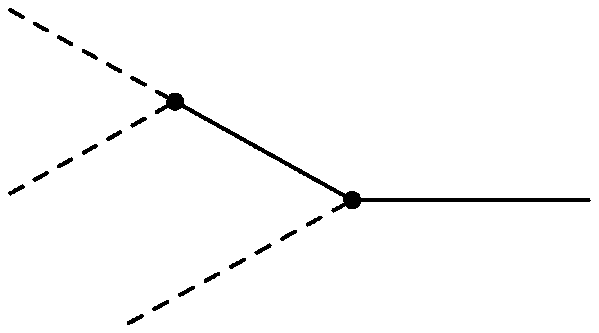}  
  \end{minipage}
  \caption{Vector-meson dominance for the main decay channels of the $\omega$ meson.
}
  \label{fig:feyn-dec-om}
\end{figure}

\begin{figure}[h!]
  \centering
  \includegraphics[keepaspectratio,width=0.69\textwidth]{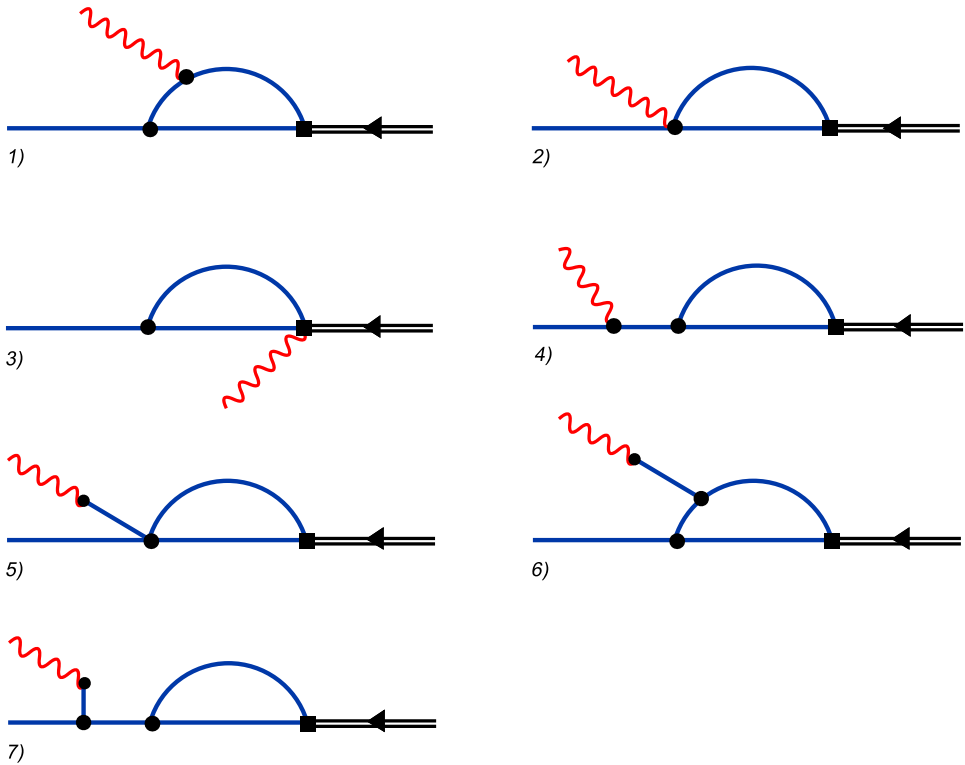}      
  \caption{No vector-meson dominance for dynamically generated states. Double lines
  denote dynamically generated states, full lines their constituents, wavy lines photons.
  See main text for details. Figure taken from \cite{lutz-leupold}.}
  \label{fig:feyn-dyn}
\end{figure}

The in-medium calculation briefly discussed in the previous section should be regarded
as a precursor to more serious considerations. Both for the understanding of the 
nature of resonances and for improved in-medium calculations, it clearly would be 
desirable to have a scheme at hand which goes beyond pure hadronic model building.
Such a scheme should allow for systematic calculations, 
i.e.\ provide a power counting such that one has a serious reason
to consider or disregard specific processes or diagrams. To operate in the
energy region of resonances such a scheme should at least contain the LLH states,
i.e.\ the pion nonet, the $\rho$ nonet, the nucleon octet, and the $\Delta$ decuplet.
For the meson sector such a scheme has been suggested recently in \cite{lutz-leupold}.
Some of its features are: Pseudoscalar and vector mesons are treated as soft.
This allows for a systematic inclusion of decays of vector mesons. It yields clear 
statements about the validity of vector-meson dominance (VMD). Finally, an interesting
aspect on the technical level is that vector mesons are represented by 
antisymmetric tensor fields. Clearly, also for one of the most interesting probes
of relativistic heavy-ion physics, the dilepton production (cf.\ the corresponding
contributions in the present proceedings) the issue of VMD is 
of central importance. A justification for the scheme proposed 
in \cite{lutz-leupold} comes from large-$N_c$ considerations, where $N_c$ denotes
the number of colors \cite{lutz-leupold}. We note in passing that an alternative 
justification emerges from the assumption of
vector mesons as dormant Goldstone bosons \cite{caldi}. Treating
both vector and pseudoscalar states as soft essentially leads to the same 
counting rules. In addition, it strongly suggests the use
of the antisymmetric tensor fields. 
We pick out two examples related to VMD for vacuum processes. According to the scheme
presented in \cite{lutz-leupold} both main decay channels of the $\omega$ meson,
the three-pion as well as the $\pi^0$-$\gamma$ decay,
are governed in leading order by VMD. This is visualized in Fig.\ \ref{fig:feyn-dec-om}.
Consequently, one can use e.g.\ the decay $\omega \to \pi^0 \gamma$ to fix the 
required coupling constant (the $\omega$-$\rho$-$\pi$ coupling) and obtain a 
prediction for the three-pion decay \cite{leupold-lutz}. One gets in that way
$\Gamma_{\omega\to 3\pi} = {7.3} \,$MeV, which is in excellent agreement
with the experimental 
value $\Gamma_{\omega\to 3\pi}^{\rm exp} = ({7.57} \pm 0.13) \,$MeV. 
One the other hand, the scheme of \cite{lutz-leupold} does {\em not} yield
VMD for every conceivable process. One counter example are the multipole moments
of the vector mesons \cite{lutz-leupold}. Another example, which we discuss now in some
detail, concerns again the dynamically generated axial-vector states. 
Here VMD does not hold as can be seen diagrammatically in Fig.\ \ref{fig:feyn-dyn}. 
The VMD process is depicted by diagram 5) of Fig.\ \ref{fig:feyn-dyn}. It relates
two decay processes to each other:
the decay of the axial-vector state into its constituents (vector and pseudoscalar
meson) and the decay into photon and pseudoscalar meson. However, also the
other diagrams shown in Fig.\ \ref{fig:feyn-dyn} contribute sizably to the
radiative decays of the axial-vector ``molecules'' \cite{lutz-leupold}.
In particular, the processes where the photon couples to the constituents
of the ``molecule'' turn out to be important, the contributions of type 1) 
in Fig.\ \ref{fig:feyn-dyn}. 
The finding that VMD does not work for the radiative decays of dynamically generated 
states is not restricted to the axial-vector mesons. It also applies e.g.\ to the
baryon resonances which play an important role for the description of the
in-medium dilepton production (see e.g.\ \cite{post} and references therein).
The absence of VMD does, of course, not imply
that the interaction of dynamically generated states with real or virtual photons
cannot be calculated. Quite on the contrary, the scheme of \cite{lutz-leupold}
(once extended to baryons)
provides a systematic framework to determine these processes. The necessary
input involves the electromagnetic moments of the constituents of the dynamically
generated states, i.e.\ of the LLH states. While these moments are well determined
for the nucleon, it is not so easy to get reliable estimates for the vector-meson
and the $\Delta$-decuplet states. For example, the vector mesons in general have
non-vanishing dipole and quadrupole moments. 
Here lattice QCD might help in the future.
Otherwise one has to determine these couplings from fits to data.

\section{Towards self-consistent in-medium calculations}

For relativistic heavy-ion physics concerning e.g.\ the production of particles 
like dileptons, hadronic in-medium calculations serve as an input 
to fireball-model or hydrodynamical calculations, but also yield in-medium cross sections
relevant for transport approaches. Besides the description of existing data and
predictions for upcoming experiments like e.g.\ CBM at FAIR (which largely operates
in the hadronic regime), one would like to understand the relation between in-medium
modifications and chiral symmetry breaking and restoration. 
To make further progress in that area on the theory
side requires several ingredients. In the following we highlight two of them:
First, a systematic framework instead of pure model building for the vacuum input and,
second, a self-consistent in-medium scheme which allows to go beyond a 
low-density expansion 
to account for the mutual interactions between the constituents of a strongly 
interacting system.
Concerning the elementary (vacuum) input 
one should incorporate the ideas of effective field theory to get from pure hadronic 
model building towards systematic approaches. The latter can assess quantitatively the 
intrinsic uncertainties and justify the neglect or incorporation of processes/diagrams.
We have discussed the development of such a scheme in the previous sections. 
At the hadronic level the elementary relevant degrees of freedom are (at least) the 
lowest-lying hadron states, LLH states. In that scheme the chiral partners of
the LLH states are not elementary at the hadronic level, but generated
from coupled-channel dynamics. In that way, one already incorporates many resonances
relevant for in-medium physics \cite{post}. Of course, there remain some states which are
not at all related to the LLH states by chiral transformations, in particular
the negative-parity mesons and the positive-parity baryons. Whether these states can 
also be generated dynamically, as advocated by the hadrogenesis 
conjecture \cite{lutz-Delta,lutz-a1,lutz-leupold}, remains to be seen \cite{lutz-FW}.
It should be clear that this scheme offers a deep relation between in-medium
physics and chiral symmetry breaking as discussed in Section \ref{sec:chirest}. The 
in-medium changes of the dynamically generated chiral partners of the LLH states
cannot be decoupled from the aspect of chiral symmetry since their shear existence
is caused by chiral dynamics. One connection to experiment are electromagnetic
observables. On the elementary level (vacuum) they serve as a diagnostic probe
for the intrinsic structure of the molecule-like states. In the context of
relativistic heavy-ion physics dileptons allow to study the in-medium properties of
hadrons. On the one hand, this concerns, of course, the vector mesons. Here 
vector-meson dominance (VMD) comes into play. As pointed out in Section \ref{sec:vmd},
the systematic scheme developed in \cite{lutz-leupold} provides clear predictions
where VMD holds in leading order and where it does not. On the other hand, in a strongly
interacting system the vector mesons in turn involve other resonances which then
also become important for the dilepton production (see e.g.\ \cite{post} and references
therein).
Finally we shall briefly comment on the required self-consistent in-medium framework.
The simplest approach to in-medium physics is the linear-density approximation.
It already provides a formidable task since in many cases the elementary input
is not completely constrained by experiment. Here the systematic scheme discussed
above comes into play. For a given hadron the linear-density approximation already 
yields in-medium changes of its properties which are due to the fact that this hadrons
interacts with the constituents of the medium. However, also the other hadrons
might change their properties as a response to the change of properties of the
originally considered hadron (``changes induce changes''). This back reaction
is not accounted for in a linear-density approximation. The mutual back reactions
should be determined in a self-consistent framework. 
Recently it has been suggested in \cite{leupold-Phi} how to
overcome some obstacles of such schemes concerning intrinsic symmetries 
like chiral symmetry or current conservation. We note in passing that one
key is the use of antisymmetric tensor fields to represent vector mesons. The 
systematic framework for the vacuum input discussed above was anyway designed in that
manner. Hence, self-consistent in-medium calculations with a systematic
vacuum input are required and possible \cite{leupold-Phi,moeller}. 
One task will be to check the validity of the melting scenario 
for chiral restoration as proposed in Section \ref{sec:chirest}.

{\bf Acknowledgments:} S.L.\ acknowledges support of the DFG to make his stay at
the International School of Nuclear Physics in Erice possible.
The work of M.W.\ has been supported by the BMBF.

\end{document}